\documentclass[twocolumn,showpacs,preprintnumbers,aps,pra,floatfix]{revtex4}

\usepackage{latexsym}
\usepackage{graphicx}
\usepackage{amsmath}
\usepackage{amssymb}
\usepackage[dvips]{color}

\begin{document}

\title{Spontaneous Emission in ultra-cold spin-polarised anisotropic
  Fermi Seas}

\author{Brian O'Sullivan} \email{bosullivan@phys.ucc.ie}
\author{Thomas Busch} 

\affiliation{ Department of Physics, National University of Ireland,
  UCC, Cork, Republic of Ireland}

\date{\today}

\begin{abstract}We examine and explain the spatial emission patterns
  of ultracold excited fermions in anisotropic trapping potentials in
  the presence of a spin polarised Fermi sea of ground state
  atoms. Due to the Pauli principle, the Fermi sea modifies the
  available phase space for the recoiling atom and thereby modifies
  its decay rate and the probability of the emitted photon's
  direction.  We show that the spatial anisotropies are due to an
  intricate interplay between Fermi energies and degeneracy values of
  specific energy levels and identify a regime in which the emission
  will become completely directional. Our results are relevant for
  recent advances in trapping and manipulating cold fermionic samples
  experimentally and give an example of a conceptually new idea for a
  directional photon source.
\end{abstract}

\pacs{05.30.-d,67.85.Lm}

\maketitle

\section{Introduction}
\label{sec:intro}

Cold samples of neutral fermionic atoms have become an important
test-bed for a large number of interesting phenomena in many particle
physics \cite{Giorgini:07}. Since the first realisation of quantum
degeneracy using two spin components of $^{40}K$ \cite{DeMarco:99},
the field has moved quickly from fundamental quantum statistical
experiments \cite{Truscott:01} into other areas such as BEC-BCS
transitions \cite{Ketterle:08}, solid state physics \cite{Bloch:08}
and even quark-gluon physics \cite{Shuryak:05}.

A particularly impressive achievement has been the realisation that by
using a Feshbach resonance it is possible to incite single atoms to
pair \cite{Bruun:04,Kohler:06}. Depending on which side of the
resonance the experiments are carried out, these pairs then either
represent bosonic molecules, which in turn form a Bose-Einstein
condensate, or Cooper-pairs, which lead to the formation of a BCS
state. By sweeping across a Feshbach resonance one can therefore
explore the BEC-BCS crossover regime, which was until recently not
experimentally available \cite{Ketterle:08}.

Besides molecular or Cooper pair physics, mono-atomic gases have also
shown a large potential for demonstrating new and exciting
physics. While the most dramatic consequence of the antisymmetry
condition on the wave-function of identical fermions is the formation
of the Fermi sea at low temperatures, other effects have been
predicted and been observed. Among them are the modification of the
scattering properties of two atoms, which leads to a reduced
efficiency of evaporative cooling \cite{Ferrari:99,DeMarco:01},
narrowing of the line-width of light propagating through the gas
\cite{Ruostekoski:98,Javanainen:99} and the suppression of
off-resonant light scattering \cite{DeMarco:98,Gorlitz:01}.

The inhibition of spontaneous emission in the presence of a ground
state Fermi sea is another fundamental prediction which results
directly from the Pauli principle
\cite{Gorlitz:01,Helmerson:90,Busch:98}. In it a degenerate Fermi sea
of a spin polarised gas forms the environment for a single, excited
atom of the same kind. Due to the Pauli principle, the Fermi sea
effectively blocks out a large amount of the phase space that would
otherwise be available to the excited atom after a de-excitation
transition. This leads to a modification of the emission properties of
the excited atom and the details of the effect are determined by the
size of the Fermi sea, the systems temperature and the anisotropy of
the trap \cite{Busch:98}. The influence on the lifetime of the excited
atom has been recently exhaustively investigated \cite{Busch:98} and
the effect was shown to be an atom-optical analogue of well known
effects in cavity QED \cite{Purcell:46}.

In this work we will investigate the influence of this Pauli-blocking
effect on the spatial distribution of the emission spectrum of a
single atom in an anisotropic trap. The fact that the emission
spectrum becomes anisotropic was first shown in \cite{Busch:98} and a
simple explanation for this effect was given. Here we investigate the
pattern formation in detail and in particular consider highly
anisotropic traps, which can be achieved today by experimentally
using, for example, atom chips or optical lattices \cite{Aubin:06}.

In Sec.~\ref{sec:Model} we will first describe the model we are using
and, in Sec.~\ref{sec:degeneracy}, derive a relation between the Fermi
energy and the number of particles in an anisotropic trap. In
Sec.~\ref{sec:structure} we describe the effect of the anisotropy on
the behaviour of the individual transition elements for spontaneous
emission and apply the results to explain the specifics of the overall
emission pattern. Finally we conclude.

\section{Model}
\label{sec:Model}

We consider an ideal gas of spin polarised fermions trapped in a
harmonic potential. All atoms are assumed to be in their internal
ground state, $|g\rangle$, so that the gas becomes quantum degenerate
at low enough temperatures and forms a perfect Fermi sea at absolute
zero. In the following we will restrict our calculations to this
limit, as in it the effects we describe are most pronounced and the
extension to finite temperatures is, while computationally
challenging, conceptually straightforward.

In addition to the Fermi sea we assume the presence of a single extra
fermion, which is distinguished from the others by being in an
internally excited state, $|e\rangle$. After some time this atom will
spontaneously emit a photon, make a transition into the ground state
and become part of the Fermi sea. As all atoms are assumed to be spin
polarised, the Pauli principle demands that the new ground state atom
has to join the Fermi sea with an energy larger than the Fermi
energy. This is an energetically very unfavourable process and the
presence of the Fermi sea will therefore lead to an inhibition of the
spontaneous emission rate with respect to the case of a free space
particle \cite{Helmerson:90,Busch:98}.

In the following we will denote the spontaneous emission rate of
photons along the direction $\Omega$ and into the solid angle $d
\Omega$ in the presence of \textit{N} ground-state fermions by
$\Gamma(\Omega)\,d\Omega$ and compare it to the free case ($N=0$)
which we denote by $\Gamma_0(\Omega) d\Omega$. Using Fermi's golden
rule we can express the excited atom's decay rate as
\begin{equation}
  \label{eq:modT}
   \frac{\Gamma(\Omega)}{\Gamma_0(\Omega)}
               = \sum_{\vec{n},\vec{m}=0}^{\infty}
                 P_m(1-F_n) |\langle\vec{n}|
                 e^{-i\vec{k}(\Omega)\cdot\hat{\vec{r}}} |\vec{m}\rangle|^2\;,
\end{equation}
where $F_n=(e^{(\hbar \omega/k_{B}T)(\vec{\lambda} \cdot
  \vec{n})}+1)^{-1}$ is the Fermi-Dirac distribution function and
$(1-F_n)$ is the probability that an energy level $\vert n \rangle$ of
the harmonic trap is unoccupied. $P_{m}=P_0
e^{-(\hbar\omega/k_BT)\vec{\lambda}\cdot\vec{m}}$ is the Boltzmann
distribution function describing the single excited fermion in state
$\vert m \rangle$ of the harmonic trap, which in turn is assumed to
have the frequencies
$(\omega_x,\omega_y,\omega_z)\equiv\omega\vec{\lambda}$. If we
restrict eq.~(\ref{eq:modT}) to zero temperature the Fermi-Dirac
distribution function becomes a step function and hence only states
with an energy greater than the Fermi energy have a finite value for
$1-F_n$. Similarly, the excited fermion will occupy the ground state
of the harmonic trap, $|m\rangle=|0\rangle$, and eq.~(\ref{eq:modT})
simplifies to
\begin{equation}
\label{eq:FGR}
 M_f(\Omega)=\frac{\Gamma(\Omega)}{\Gamma_0(\Omega)}
            =\sum_{n=n_F+1}^{\infty} 
             |\langle\vec{n}|e^{-i\vec{k}(\Omega)\cdot\hat{\vec{r}}}|0\rangle|^2\;,
\end{equation}
where $n_F$ represents the Fermi shell. The inhibition of spontaneous
emission that results from equation (\ref{eq:modT}) has recently been
investigated \cite{Busch:98} and here we extend this work by
presenting a thorough and detailed investigation into the spatial
emission patterns that result from it. This is of interest due to the
new parameter ranges which have become experimentally available in
recent years. These include, in particular, highly anisotropic
traps. While the appearance of a fine structure in the emission
patterns was already mentioned in \cite{Busch:98}, here we derive the
framework for its explanation.

We assume that the harmonic oscillator potential has the following
standard form
\begin{equation}
  \label{eq:HarmonicPotential}
  V=\frac{M \omega^2}{2}(\vec{\lambda}\cdot\vec{r})^2
   =\frac{M \omega^2}{2}(\lambda_x^2x^2+\lambda_y^2y^2+\lambda_z^2 z^2)\;,
\end{equation}
where $M$ is the mass of the particle and the values of
$\lambda_{x,y,z}$ determine the degree of anisotropy in the different
directions. For numerical simplicity we will only deal with values of
$\lambda \ge 1$ and restrict ourselves to two types of anisotropic
trapping potentials in which two of the axes have identical strength,
\begin{subequations}
  \label{eq:trapshapes}
\begin{align}
  \lambda_x&=\lambda_y=1 \ \text{and} \
   \lambda_z=\lambda,\qquad\text{pancake shape}\;,\\
  \lambda_x&=\lambda_y=\lambda \ \text{and} \
   \lambda_z=1,\qquad\text{cigar shape}\;.
\end{align}
\end{subequations}
Due to the symmetries of the pancake and cigar shaped traps about the
tight and the soft axes, respectively, eq.~(\ref{eq:FGR}) can be
simplified and expressed in terms of incomplete gamma functions for
both trap shapes
\begin{equation}
  \label{eq:modfac}
   M_f(\theta) = \frac{\gamma(n_F+1,\beta)}{\Gamma(n_F+1)}
                 +e^{-\beta}\sum_{n=0}^{n_F}\frac{\beta^{n}}{n!}
                 \frac{\gamma(\lfloor\frac{n_F-n}{\lambda}\rfloor+1,\frac{\alpha}{\lambda})}
                 {\Gamma(\lfloor\frac{n_F-n}{\lambda}\rfloor+1)}\;.
\end{equation}
Here $\lfloor$x$\rfloor$ denotes the largest integer less than or
equal to x and $n_F$ is the quantum number of the Fermi shell, which
we will describe in more detail later. For brevity we have defined
\begin{subequations}
  \label{eq:trapabrev}
\begin{align}
  \alpha&=\eta^2\cos^2\theta,
  \quad\beta=\eta^2\sin^2\theta\qquad\text{pancake shape}\;,\\
  \alpha&=\eta^2\sin^2\theta,
  \quad\beta=\eta^2\cos^2\theta\qquad\text{cigar shape}\;,
\end{align}
\end{subequations}
in which $\eta$ represents the Lamb-Dicke parameter. One can
immediately see that the angular distributions for the pancake and the
cigar shaped trap can be obtained from each other by a simple $\pi/2$
rotation, as one would expect. We will make use of this fact when
discussing emission patterns in Sec.~\ref{sec:structure}.

\section{Degeneracies}
\label{sec:degeneracy}

Let us first discuss the relationships between the different
parameters characterising a Fermi-sea in an anisotropic trap. Since
the degeneracies of states with equal energy are a function of the
trapping frequencies in the different directions, the relationship
between the Fermi energy and particle number is not as straightforward
as in the well-known isotropic case.

The eigenenergies of the harmonic potential in
eq.~\eqref{eq:HarmonicPotential} are given by
\begin{subequations}
  \begin{align}
    \label{eq:EPan}
    E_{n_\text{p}}&=
      \left(n_\text{p}+\left(\frac{\lambda}{2}+1\right)\right)
      \hbar\omega,\\
    \label{eq:ECig}
    E_{n_\text{c}}&=
      \left(n_\text{c}+\left(\lambda+\frac{1}{2}\right)\right)
      \hbar\omega,
  \end{align}
\end{subequations}
where we have defined the shell quantum numbers of the pancake and
cigar shaped harmonic traps as $n_\text{p} = n_x + n_y + \lambda n_z$
and $n_\text{c}= \lambda n_x +\lambda n_y+ n_z$, respectively. As
usual, $n_x, n_y$ and $n_z$ refer to the integer quantum numbers of
the harmonic oscillator.

As the aspect ratio of a trapping potential is increased the resulting
energy levels typically have a reduced degeneracy relative to the
isotropic case, $\lambda=1$ \cite{Cohen-Tannoudji}. For the purposes
of this work, and without loss of generality, we will consider only
integer values of $\lambda$, allowing us in turn to restrict ourselves
to integer values for $n_\text{p,c}$. Therefore we can write the
degeneracy for states with fixed energy as
\begin{subequations}
  \begin{align}
   \label{eq:DegenPan}
   g_{n_\text{p}}&=\frac{1}{2}(\tilde n_\text{p}+1)
                  (2n_\text{p}-\lambda\tilde n_\text{p}+2)\;, \\
   \label{eq:DegenCig}
   g_{n_\text{c}}&=\frac{1}{2}(\tilde n_\text{c}+1)(\tilde n_\text{c}+2)\;,
  \end{align}
\end{subequations}
where here and in the following all quantities carrying a tilde take
the value of the quantity without the tilde divided by $\lambda$ and
rounded down to the nearest integer, i.e. $\tilde x = \lfloor
x/\lambda \rfloor$. Consequently, the total number of quantum states
with an energy equal to and smaller than $E_{n_F}$ is then given by
the sum,
\begin{equation}
  \label{eq:sum}
  S=\sum_{n=0}^{n_F} g_n\;,
\end{equation}
which can be calculated to be given by
\begin{subequations}
  \begin{eqnarray}
  \label{eq:sumpan}
      S_\text{p}&=&
        \frac{1}{6}(\tilde n_F+1)(2n_F-\lambda\tilde n_F+2)\nonumber\\ 
        &&\left(\frac{3}{2}n_F-\frac{3}{4}\lambda\tilde{n_F}
                +\frac{\lambda^2\tilde n_F(2+\tilde n_F)}
                      {8+8n_F-4\lambda\tilde n_F}+3\right),  \\
  \label{eq:sumcig}
      S_\text{c}&=&
       \frac{1}{6}(\tilde n_F+1)(\tilde n_F+2)(3n_F-2\tilde n_F\lambda+3).
  \end{eqnarray}
\end{subequations}
In our model we assume a spin polarised gas in which each oscillator
state is filled with one fermion only.  Eqs.~\eqref{eq:sum} therefore
determine the number of particles confined for a given Fermi energy
$E_F = n_F \hbar \omega +E_G$, where $E_G$ is the ground state energy
of the potential.

\section{Emission Patterns in Anisotropic Traps}
\label{sec:structure}

\subsection{Emission Probabilities.}
\label{sec:enocc}

To understand the emission patterns later on, let us first have a
brief look at the emission probabilities of an excited atom in an
anisotropic trap. In the presence of an anisotropic Fermi sea the rate
of spontaneous emission along a specific direction is determined by
three parameters: (1) the number of ground state atoms, (2) the
degeneracy of the available states and (3) the Lamb-Dicke parameter
$\eta = \sqrt{E_R/\hbar \omega}$. The latter determines the range of
accessible states and is given by the ratio between the recoil energy,
$E_R=\hbar^2 k_0^2/2M$, and the trapping strength, $\hbar
\omega_{x,y,z}$, in the different directions. Here $k_0$ is the wave
vector corresponding to the transition $|e\rangle \rightarrow
|g\rangle$.

\begin{figure}[tb]
  \includegraphics[width=\linewidth,clip]{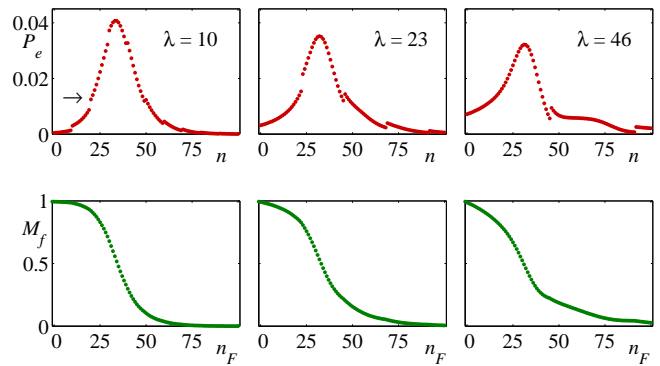}
  \caption{Top row: emission probability, $P_e$, into individual
    shells in a pancake shaped trap for $\eta^2 = 36$ and $\lambda =
    10,23,46$. The arrow indicates the $n=20$ energy level of the
    harmonic trap, which is referred to in the text.  Bottom row:
    decay rate of the excited particle, $M_f$, for the same parameters
    as above.}
\label{fig:pancakeER}
\end{figure}

\begin{figure}[tb]
  \includegraphics[width=\linewidth,clip]{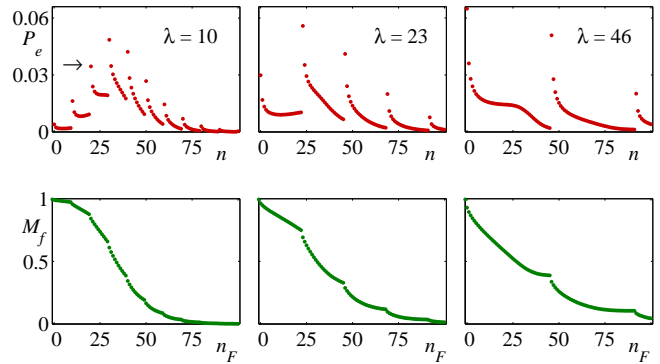}
  \caption{Top row: emission probability, $P_e$, into individual
    shells in a cigar shaped trap for $\eta^2 = 36$ and $\lambda =
    10,23,46$. The arrow indicates the $n=20$ energy level of the
    harmonic trap, which is referred to in the text. Bottom row: decay
    rate of the excited particle, $M_f$, for the same parameters as
    above.}
\label{fig:cigarER}
\end{figure}

In this section we will focus on the influence of the degeneracies of
the available states and therefore on the anisotropy of the trap. Let
us do this by examining the matrix elements for individual transitions
from the ground cm-state of the excited atom to a single final state,
$|n\rangle$,
\begin{equation}
\label{eq:matrixelement}
  P_e(n) = |\langle n |e^{i k x}|0 \rangle|^2\;.
\end{equation}
It is well known that for an isotropic trap this distribution is
Gaussian in shape and centered around an energy level $n =
\eta^2$. The effects introduced by an anisotropy are significant and
can be clearly seen in the graphs in the upper rows of
Figs.~\ref{fig:pancakeER} and \ref{fig:cigarER}, where we show $P_e$
for a pancake and a cigar shaped trap, respectively, for increasing
values of the anisotropy, $\lambda =$ 10, 23 and 46.  The most obvious
feature in both situations is the appearance of a $\lambda$-dependent
discontinuity in the distribution, which is more pronounced in the
cigar shaped setting.

To explain this behaviour, let us first intuitively argue its
existence.  When an internally excited atom which is trapped in the
ground state of an empty isotropic harmonic trap decays, the
probability of the photon being emitted is the same in all directions.
This is rather easy to understand as in this situation the density of
states is identical in all directions. However, for the anisotropic
trap the situation is different. As the aspect ratio is increased, the
degeneracy of any energy level will either decrease or remain the
same. Therefore, up to a specific shell the number of quantum states,
given by eqs.~\eqref{eq:sum}, will be reduced and, as a result, the
density of states in the different directions changes. Since the
recoil of the de-excited fermion to a certain quantum state and the
direction of the emitted photon are directly related, it seems rather
surprising that for the free case this emission remains isotropic
irrespective of the diminishing number of quantum states.  However, it
is exactly the modified distribution shown in the upper rows in
Figs.~\ref{fig:pancakeER} and \ref{fig:cigarER} that makes this
phenomenon possible.

\begin{figure}[tb]
\includegraphics[width=\linewidth,clip]{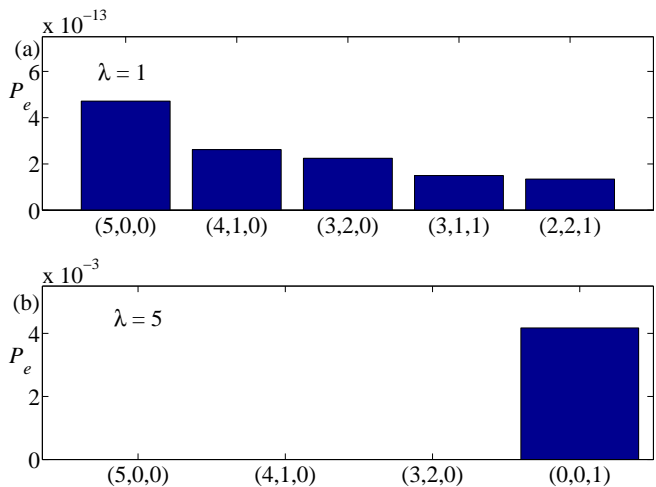}
\caption{(a) Emission probability into individual states within the
  shell $n=5$ in an isotropic trap. The triplets represent $(n_x, n_y,
  n_z)$ and all permutations of each triplet have the same
  probability. The Lamb-Dicke factor is $\eta=5$, however changing
  this value only scales all values. (b) Emission probability into
  individual states within the shell $n=5$ in a pancake trap with
  $\lambda=5$. The values for the three states on the left are not
  visible on this scale.}
\label{fig:shellstates}
\end{figure}

To gain more insight into the source of the discontinuities let us
consider the emission probability into specific states within a
degenerate shell $n$ of an isotropic and a pancake shaped
$(\lambda=5)$ trap. Fig.~\ref{fig:shellstates} shows
$P_e(n_x,n_y,n_z)$ for a fixed shell, in which all combinations of the
triplet (in both (a) and (b)) of quantum numbers adds up to $n=5$. It
can be seen that in general states which include ground state
excitations have a higher probability for occupation than the ones
which do not, which is due to the fact that the excited atom is
initially in its centre-of-mass ground state.

When we move from the isotropic to the anisotropic setting it is
therefore clear that whenever the value of an energy shell, $n$,
reaches an integer multiple of the anisotropy parameter, the shell
contains a state with two ground state excitations. As these states
have a higher probability of occupation (see
Fig.~\ref{fig:shellstates}(b)) the overall emission probability into
this energy shell is increased, leading to the observed discontinuous
jump. As an example let us consider a pancake shaped trap with an
aspect ratio of $\lambda=10$. For the shell $n = 19$ (indicated by the
arrow in Figs.~\ref{fig:pancakeER}) the degenerate states are
$(n_x+n_y,n_z)=(19,0)$ and $(9,1)$, whereas for the $n=20$ energy
level states are $(n_x+n_y,n_z)=(20,0), (10,1)$ and $(0,2)$.  The {\sl
  extra} $(0,2)$ state is the dominant contributor and its appearance
responsible for the discontinuous increase in the emission
probability. For the cigar trap this effect is even more pronounced as
there are two tight directions and in the example above the states
$(n_x,n_y,n_z)=(2,0,0)$ and $(0,2,0)$ become both available.

For completeness we show the integrated emission probability for
increasing particle number (i.e. increasing Fermi energy or Fermi
level) and different anisotropies in bottom rows of
Figs.~\ref{fig:pancakeER} and \ref{fig:cigarER}. Fermi inhibition is
absent for the empty trap ($M_f=1$), shown for $n_F=-1$, slowly
increases for $n_F \ge 0 $ and accelerates for $n_F \sim \eta^2$. The
discontinuity in the variable $P_e(n)$ translates clearly into
non-smooth kinks in this distribution.

\subsection{Emission along Tight and Soft Axes.}
\label{sec:tightsoft}

The fact that the presence of an anisotropic Fermi sea will lead to
anisotropic emission patters was already noted in \cite{Busch:98} and
in the following we will develop a detailed understanding of the
directional features. Since for the pancake as well as for the cigar
shaped trap the emission is isotropic around their respective symmetry
axis, $(0,\pi)$, we can treat both geometries in a quasi 2D picture.
It is then immediately clear that the results for both settings will
be related by a simple $\pi/2$ rotation (due to our definition of
$\lambda\ge1$).

\begin{figure}
  \includegraphics[width=\linewidth,clip]{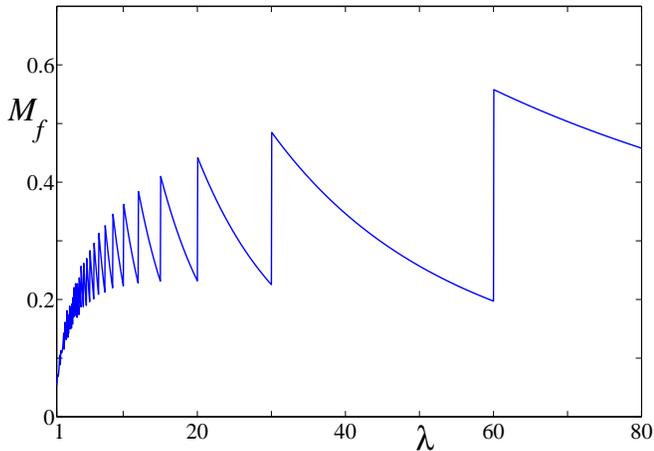}
  \caption{$M_f$ along the tight axis at $T=0$. $\eta^2 = 49$,
    $n_F=60$. Note that we use a continuous distribution of $\lambda$
    for this graph.}
  \label{fig:theta0}
\end{figure}

Let us first look at the emission along the principal axes of the
anisotropic trap in the tight and the soft direction. Choosing the
tight direction in the pancake (cigar) shaped trap along $\theta=0$
($\theta=\pi/2$) the modification factor in eq.~\eqref{eq:modfac}
simplifies to
\begin{equation}
 \label{PAemission}
 M_f=\frac{\gamma(\tilde n_F+1,\frac{\eta^2}{\lambda})}
 {\Gamma(\tilde n_F+1)}\;.
\end{equation}
The behaviour of this equation with increasing anisotropy is shown in
Fig.~\ref{fig:theta0} for a system with $n_F=60$.  The most obvious
feature of the plot is a series of sawtooth-like
discontinuities. Careful examination shows that $n_F$ of these exist
and they appear whenever the value of the aspect ratio, $\lambda$,
increases beyond the values of $\frac{n_F}{m}$, ($m =
1,2,\dots,n_F$). The increase in emission probability for values just
after this point is due to the availability of an extra free state
with a lower tight excitation just outside the Fermi edge. For
example, in the pancake trap, when one moves from $\lambda=30$ to
$\lambda>30$ the state $(n_x+n_y,n_z)=(0,2)$ emerges from the Fermi
sea for $n_F=60$. As discussed in Section~\ref{sec:enocc}, this state
has a high probability to be emitted into as it contains ground state
excitations in the soft direction, hence the large increase in the
decay rate. By increasing $\lambda$ further this state moves away from
the Fermi edge, and the emission probability decreases until the next
state with a lower tight excitation, emerges from the Fermi sea. For
values of $\lambda>n_F$ no more discontinuities appear since the Fermi
sea only occupies energy states with ground state excitations in the
tight direction. Emission along the soft direction can be calculated
from eq.~\eqref{PAemission} by taking $\lambda=1$ and the decay rate
along this direction is determined exclusively by the Fermi shell
$n_F$ and the value of the Lamb-Dicke parameter $\eta$.

\begin{figure}
  \includegraphics[width=\linewidth,clip]{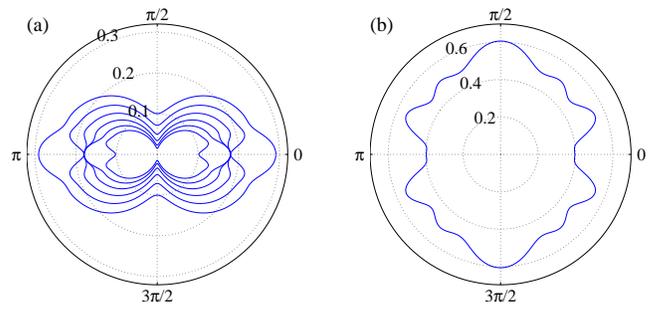}
  \caption{(a) $M_f(\theta)$ in a pancake shaped trap at
    $T=0$. $\eta^2 = 25$, $\lambda=4$ with $n_F = 31$ (outermost) to
    $n_F=36$ (innermost). (b) $M_f (\theta)$ in a pancake shaped trap
    with $\lambda=11$, $\eta^2=25$ and $n_F=23$. }
  \label{fig:degenplot}
\end{figure}

Considering a fixed value of the aspect ratio $\lambda$ in either
anisotropic trapping potential and changing $n_F$ one notices a
degeneracy in the emission probability in the tight direction, shown
in Fig.~\ref{fig:degenplot}(a). This behaviour was already mentioned
in \cite{Busch:98} and we can see from eq.~(\ref{PAemission}) that it
stems from the fact that $\tilde n_F$ only changes its value in steps
of $\lambda$. An increase in the value of $\tilde n_F$ coincides with
the Fermi sea occupying a state with a higher tight excitation (and
ground state soft excitations), leading to a decrease in the decay
rate along the tight direction. For example, when moving from $n_F =
35$ to $n_F = 36$ the state $(n_x+n_y,n_z)=(0,9)$ becomes occupied by
the Fermi sea, producing the discontinuous reduction of the decay
rate (see Fig.~\ref{fig:degenplot}(a)).

\subsection{Fine Structure}
\label{sec:finestruct}

\begin{figure}
\includegraphics[width=\linewidth,clip]{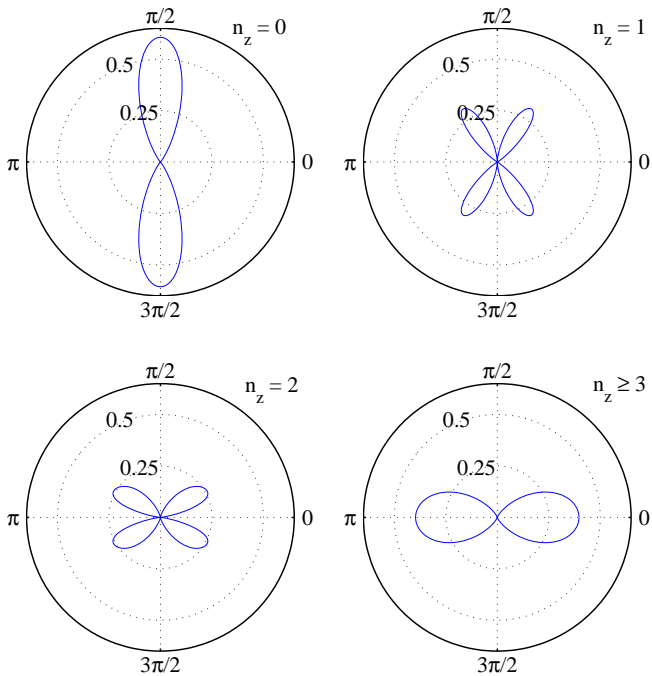}
\caption{$M_f(\theta,n_z)$ for a pancake shaped trap with, $\lambda =
  11$ $\eta^2 = 25$ and $n_F = 23$. In the four graphs the decay is
  only allowed into quantum states of the harmonic trap with $n_z =$
  0, 1, 2, and $n_z \geq$ 3, respectively.}
\label{fig:finestructure}
\end{figure}

The emission spectrum between the principal axes is characterised by
the appearance of a fine structure (see Fig.~\ref{fig:degenplot}(b)),
which exists for a wide range of parameters.  The first hint to
understanding the origins of the visible extrema comes from noticing
that the number of maxima between the soft and tight axes is related
to the number of excitations in the tight direction that are occupied
by the Fermi sea, $\tilde n_F$.  To show this relation let us consider
the emission probability into shells with a fixed value for $n_z$ in a
pancake shaped trap
\begin{equation}
  \label{eq:tem}
  M_f(\theta,n_z)=e^{-\frac{\alpha}{\lambda}} 
  \frac{\left(\frac{\eta^2}{\lambda}\right)^{n_z}}{n_z!}
  \frac{\gamma(\textrm{max}(0,n_F-\lambda n_z +1),\beta)}
       {\Gamma(\textrm{max}(0,n_F-\lambda n_z +1))}
\end{equation}
with the definition of $\alpha$ and $\beta$ given in
eqs.~(\ref{eq:trapabrev}).  As a specific example we show in
Fig.~\ref{fig:degenplot}(b) a gas with $n_F=23$, $\eta^2=25$ and
$\lambda=11$. In this case we find $\tilde n_F = 2$ maxima in the
$\pi/2$ arc between the tight to the soft axis. Comparing this
emission pattern to the results from eq.~\eqref{eq:tem}, one can see
(Fig.~\ref{fig:finestructure}) that each isolated contribution from a
transition into a state with a fixed value of $n_z$ is responsible for
one of the maxima. For values of $n_z > \tilde n_F$ the emission is
predominantly into the tight direction, therefore originating from
transitions into states for which both ground state excitations in the
soft direction are available. Similarly, when restricting the
recoiling atom to occupying states with a ground state excitation in
the tight direction, $n_z=0$, the emission is mainly focussed around
small angles about the soft axis. The intermediate excitations,
$n_z=1,2$, make up the two intermediate ripples between the principle
axes and summing up the contributions to the photon emission of all
four plots in Fig.~\ref{fig:finestructure} gives the emission plot
shown in Fig.~\ref{fig:degenplot}(b). In contrast, if we calculate
eq.~\eqref{eq:tem} for an isotropic trap for different values of
$n_z$, each individual term would show a similar behaviour of having a
single maximum at a finite angle between the principle axes. However,
the sum of those will give the isotropic emission pattern which
corresponds to the decay rate being the same in all directions.

\begin{figure} 
  \includegraphics[width=\linewidth,clip]{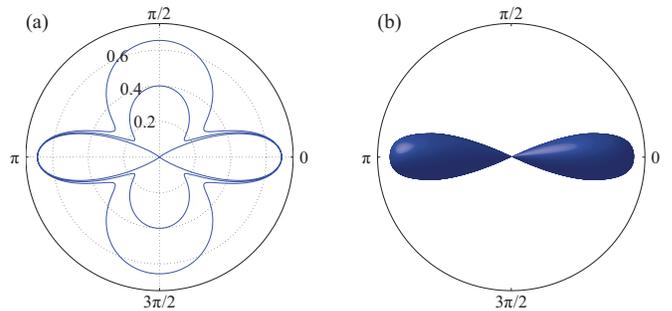}
  \caption{(a) $M_f(\theta)$ in a cigar shaped trap at T = 0 with $n_F
    = 45$ and $\eta^2 = 49$. $\lambda = 46$ (outermost), $\lambda =
    96$ (center plot) and $\lambda = \infty$ (innermost). The plot is
    symmetric through a $2 \pi$ rotation about the $(0,\pi)$ axis. (b)
    A three-dimensional illustration of the excited particles decay
    rate in a cigar trap in the large anisotropy limit.}
\label{fig:directedemission}
\end{figure}

It is now obvious that for the limit $\lambda > n_F$ the fine
structure disappears and the extrema of emission will be located
around the directions of the principal axes (see
Fig.~\ref{fig:directedemission}(a)). As $\lambda \rightarrow \infty$,
emission into the tight direction is reduced, whereas the emission in
the soft direction remains constant,
Fig.~\ref{fig:directedemission}(b). In this regime the Fermi sea
is completely confined to states with ground state excitations
in the tight direction. Therefore, it becomes easier for the
recoiling atom to access states in the soft direction due to the
diminishing density of states in the tight direction. In the limit of
$\lambda\rightarrow\infty$ the emission probability can be written as

\begin{equation}
  \label{eq:lambdainfty}
  M_f(\theta;\lambda\rightarrow\infty)=
   \frac{\gamma(n_F+1,\beta)}{\Gamma(n_F+1)}\;,
\end{equation}

and it shows that the emission probability in the tight direction has
completely vanished.

It is possible to make use of this behaviour and create a system where
photon emission become highly directional. While directional photon
emission is usually achieved by using optical cavities (and therefore
engineering the Hilbert space of the photon), this example is
complementary in that it uses a cavity (trap) for the atoms and
thereby engineers the Hilbert space of the particles. Let us stress
that it is not primarily the size of the Fermi sea that is responsible
for this effect, mearly the presence of the Fermi sea. The emission
probability of the photon can still be close to the emission probability
in free space whilst $\eta^2 \gtrsim n_F$.
(see Fig.~\ref{fig:directedemission}(a)). As the emission is symmetric
through a $2\pi$ rotation about the $(0,\pi)$ axis in the above
example, we display the 3D emission probability in
Fig.~\ref{fig:directedemission}(b). Also note that for a pancake
shaped trap this effect would correspond to emission into a well
defined plane perpendicular to the tight principal axis.

\section{Conclusion}
\label{sec:conclusion}

In this work we have given a detailed investigation into the spatial
properties of spontaneous emission of a single atom in the presence of
an anisotropic, ideal and spin polarised ultracold Fermi gas. The
demand to obey the Pauli principle leads to the formation of a
non-trivial, anisotropic emission pattern for the photon, which can be
explained by carefully examining the allowed transitions the recoiling
atom can make.

We have first calculated the relation between the Fermi energy and the
particle number and then investigated the single particle transition
matrix element, for both geometries of anisotropic traps. The change
in the density of states into the different spatial directions was
found to be accompanied by the appearance of discontinuities in the
distribution of the emission probability spectrum for different
shells. While in an isotropic trap these two effects cancel and
produce an isotropic emission spectrum, in an anisotropic trap they
lead to an intricate fine-structure in the presence of a Fermi sea.

In a next step we have managed to explain this fine-structure by
attributing the extrema to the emissions which come from the
transitions of the recoiling atom into well defined states in the
tight direction. If the aspect ratio exceeds the Fermi energy, the
fine-structure vanishes and the emission spectrum becomes smooth,
though not isotropic, again.

Finally, we have pointed out that this system can be used to create a
highly directional photon source. The effect uncovered is
complementary to the common use of optical cavities to influence a
photons direction after emission and makes use of the ability to
influence the atom's phase space. The experimental observation of
directional photon emission in anisotropic, cold, fermionic gases
would therefore be a sign of a fundamental consequence of the symmetry
of fermionic particles.

\section{Acknowledgements}

We would like to thank T.~Ramos for valuable discussions. This project
was supported by Science Foundation Ireland under project number
05/IN/I852. BOS acknowledges support from IRCSET through the Embark
Initiative RS/2006/172.

\end{document}